# Non-Canonical GC Base Pairs and Mechanochemical Cleavage of DNA

Denis A. Semyonov[1], Yury D. Nechipurenko[2]

[1]*Institute of Biophysics, Siberian Branch of Russian Academy of Science, Russia*
[2]*Engelhardt Institute of Molecular Biology, Russian Academy of Sciences, Moscow, Russia*

*When you have eliminated the impossible, whatever remains, however improbable, must be the truth.*

*Abstract. Properties of non-canonical GC Base Pairs and their relation with mechanochemical cleavage of DNA are analyzed. A hypothesis of the involvement of the Transient GC Wobble Base Pairs in the mechanisms of the mechanochemical cleavage of DNA and epigenetic mechanisms with participation of 5-methylcytosine is proposed. The hypothesis explains the increase in the frequency of the breaks of the sugar-phosphate backbone of DNA after cytosines, asymmetric character of these breaks, and an increase in the frequency of breaks in CpG after cytosine methylation.*

In the works of S.L. Grokhovsky and co-authors, a non-random character of the ultrasound-induced fragmentation of DNA was shown: the frequency of the DNA sugar-phosphate backbone breaks strongly depends on the base sequence [1-4]. Nowadays, owing to the usage of the ultrasound-induced and other type of mechanochemical cleavage of DNA in the process of the new generation sequencing (NGS), a large bulk of data confirming these findings has been accumulated [5, 6].

On the other hand, in the last decade, new data concerning the abundance and biological significance of non-canonic base pairs have been obtained. Such interactions quite often involve tautomeric or ionized forms of nucleotides [7, 8]. Transient pairs are found in the DNA double strand [9]. In this work we consider a possible role of such transient pairs in the increase of the frequency of the ultrasound-induced DNA cleavages.

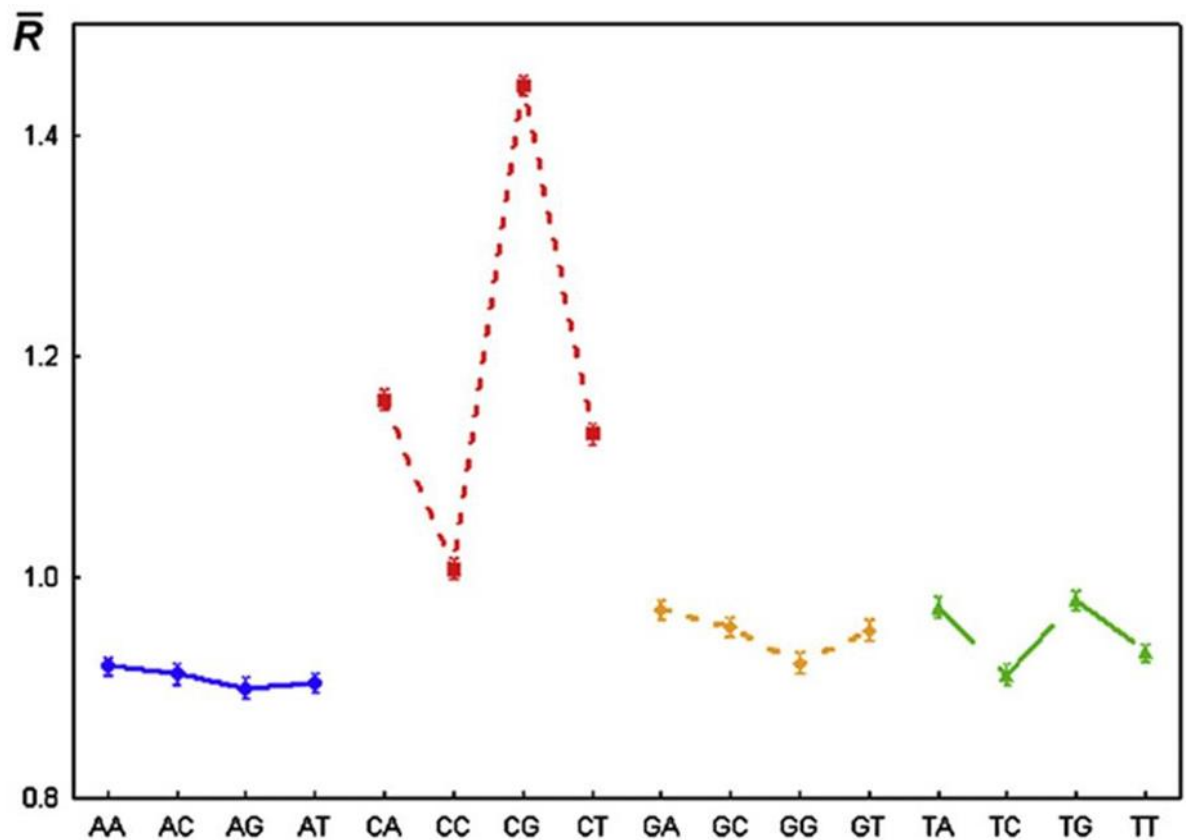

*Fig. 1. Mean values of the relative cleavage rates for all dinucleotides, and 95% confidence limits for the population mean. Symbols: ■, ▼, ◊, ○ - mean value; I - 95% confidence limits for the population mean.*

Available data on the frequency of ultrasound-induced DNA cleavages point to a non-random nature of this process. Most often the breakages occur after cytosine (deoxycytidine) and significantly depend on the nearest neighbor. Of special importance is dinucleotide CpG: the frequency of the sugar-phoshate backbone breakage in the DNA chain (in the direction 5'–3') is about one and a half as much as that of other dinucleotides not containing cytosine in the first position (Fig. 1).

It was proposed in [4] that the increase in the frequency of DNA cleavage is related with an increase of the projection on the DNA axis of the bond undergoing cleavage. The increase of the bond projection ensures for an increase of the force applied to this bond upon stretching of the sugar-phosphate backbone in the process of the DNA stretching upon the cavitation bubble collapse. The authors of [4] relate the projection increase with the N – S transition in deoxyribose linked with cytosine. Further, not attempting to reject this assumption, we will try to point to another possible reason for the observed effect.

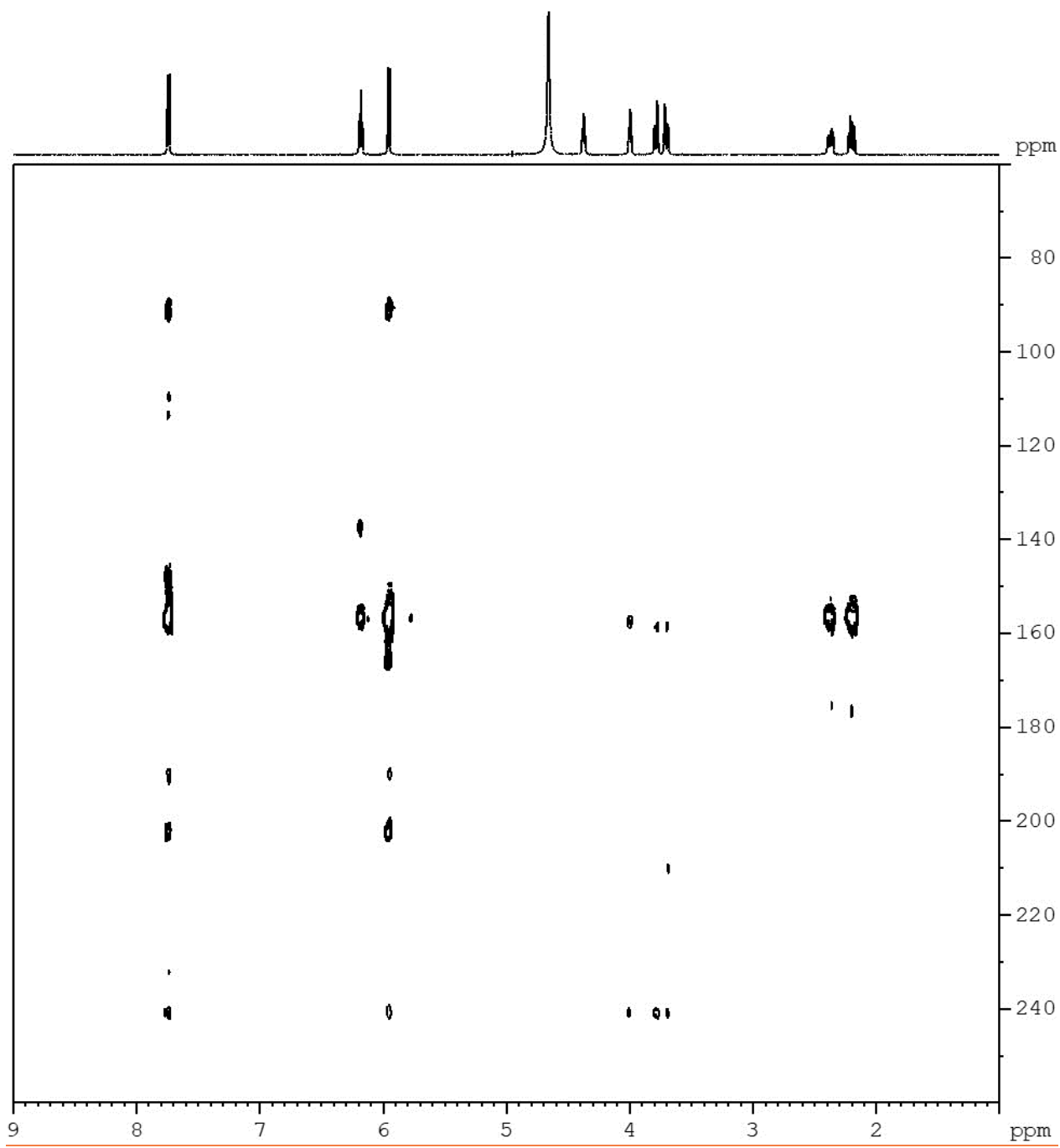

*Fig. 2. 1H15N NMR of the aqueous solution of deoxycytidine at pH 7. At least five different nitrogen atoms are detectable, which can be explained by the existence of a considerable fraction of both amino- and imino- tautomer (experimental data are provided by Dr. Ilia V. Eltsov, Novosibirsk State University).*

The cause of the cleavage frequency increase can be searched in the cytosine properties. An increased ability to create a tautomer (imino) form is an inherent feature of cytosine, and this is observed in the 1H15N NMR spectra of the deoxycytidine solution (Fig. 2). In the spectrum shown, the fraction of the protonated form at pH 7 is negligible, which is also confirmed in other sources [8]. Apparently, the ability of cytosine to form tautomeres is not harmful for replication. It should be mentioned for comparison that although in the 1H15N NMR spectrum of the uridine solution no signs of the presence of the tautomer (enol) form are found, the presence of the GU pairs with this form of uridine has been currently confirmed both by X-Ray crystallography [7, 10] and by NMR [11].

In addition to the propensity of cytosine for tautomerization, a possibility of protonation of nitrogen N3 in a weakly acidic medium should be mentioned. It is the presence of the protonated form of cytosine is thought to account for the ability of the GC pairs to form Transient GC Hoogsteen Base Pairs in the DNA Double Helix [8]. It is noteworthy that the tautomer form of cytosine is also potentially able to form such pairs (Fig. 3).

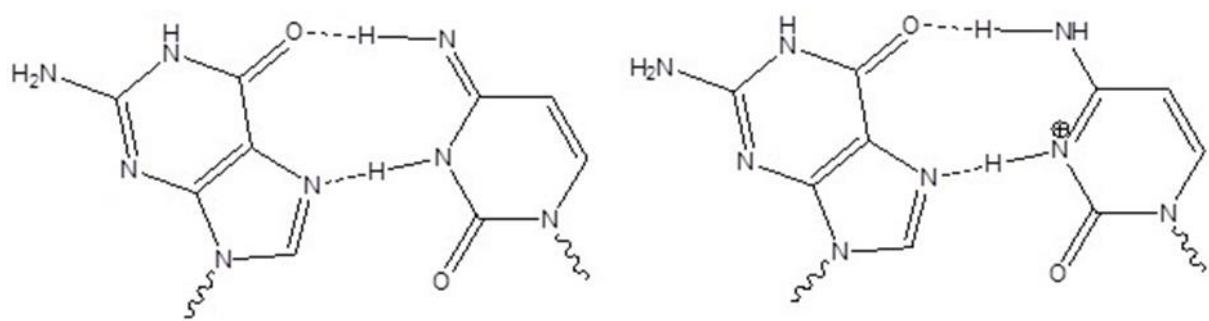

*Fig. 3. Transient GC Hoogsteen Base Pairs. GC-imino (left) and GC+ (right).*

According to the review of Al-Hashimi at al. [12], Hoogsteen Base Pairs always locally distort the geometry of the DNA double helix by forming a sharp bend in the direction to the DNA major groove . Besides, Hoogsteen Base Pairs produce a decrease of the distance between the strands of the double helix. If the estimates are made on the basis of the distance between the C1' atoms of deoxyriboses, then the DNA double helix is more compressed in the case of the pair GC Hoogsteen than for the AT Hoogsteen Base Pairs.

Local distortion of the double-helix geometry may be a physical cause of the increased frequency of the DNA cleavage induced by ultrasound. Energy required for the transition of the AT pair to the Hoogsteen Base Pair is obviously lower than of the GC pair, as breaking of only one hydrogen bond is needed. Experimental data indicate that the number of such Transient AT Hoogsteen Base Pairs in Duplex DNA must be larger than the number of GC [13]. However, an increase in the frequency of the ultrasound-induced DNA cleavage after AT pairs is not observed [1-6]. This is an essential argument against the assumption that Hoogsteen Base Pairs are responsible for the increase in the frequency of the ultrasound-induced DNA cleavage.

One more source of the possible distortions of the double-helix geometry can be Transient GC Wobble Base Pairs, which were not included earlier into the analysis of the Transient Base Pairs in Duplex DNA. GC Wobble Base Pairs also have to distort the geometry of the DNA double helix. Tautomerization or protonation of cytosine makes the existence of GC Wobble Base Pairs possible (Fig. 4). To suggest an analog of Wobble Base Pair For AT is impossible. It should be noted that this fact agrees well with a low frequency of the ultrasound-induced DNA cleavage after AT pairs.

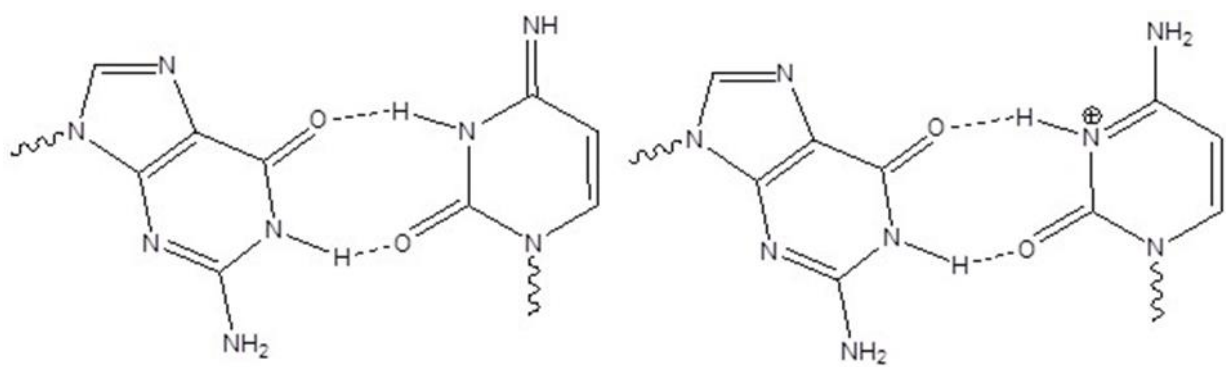

*Fig. 4. Transient GC Wobble Base Pairs. GC-imino (left) and GC+ (right).*

An analog of GC Wobble Base Pair was found in the course of investigations of N4-methoxycytosine (mo4C) in the DNA double helix [14]. N4-methoxycytosine is form imino tautomer. For this modified derivative of cytosine, the existence of two forms (mo4CG Wobble and mo4CG Watson-Crick) in a dynamic equilibrium was shown.

It is noteworthy that the information on the structure of Wobble Base Pairs in DNA can be obtained from the analysis of the data concerning the GT Wobble Base Pairs. Such pairs emerge in DNA as a result of deamination of 5-methylcytosine. RSA data on the structure of GT Base Pairs in the DNA double helix are available for at least 30 years [15], and respective NMR data, for at least 20 years [16]. The analysis of the structure of GT Wobble Base Pairs performed by both methods points to a distortion localized within one base pair [15, 16]. This distortion could result in the localization of the ultrasound-induced DNA cleavage. Of importance may be an asymmetric location of the GT Wobble Base Pair relative to the sugar-phosphate backbone of the DNA. The distance between atoms C' of deoxyriboses in the structure of the GT Base Pair in the DNA double helix coincides with the GC WC Base Pair parameters [16]. For B-DNA, angles between bonds N1\N9-C1' and vector C1'-C1' are close to 55º.

As a result of the base shift, these values considerably deviate in the case of GT Wobble Base Pair: angles between bonds N1\N9-C1' for this pair differ almost twofold (40º and 70º, respectively) [16]. Sugar-phosphate backbone should also be deformed asymmetrically and will be additionally tensed in the proximity of (near) the pyrimidine of Wobble base pair. This can explain the ultrasound-induced breakage of the bond after cytosine and retaining unaltered frequency after guanine at the middle level.

Thus, to check the significance of the GC Wobble Base Pairs, DNA with incorporated GT Base Pairs can be used for studies of ultrasound-induced cleavage of such DNA. In this case, a high frequency of breakages of the DNA Wobble pairs after thymine may serve a strong argument evidencing that an increase in the DNA cleavage frequency may be related with the formation of the GC Wobble Base Pairs.

In the works of Grokhovsky and co-authors experiments were carried out on the plasmid DNA, and the results of the NGS sequencing were analyzed for the eukaryotic DNA [5, 6]. Methylated cytosine is absent in plasmides, and a certain part of CpG dinucleotides is methylated in the eukaryotic DNA. In the eukaryotic DNA, an increase of the frequency of the DNA cleavage at the dinucleotide CpG bond is observed, and the authors of [5, 6] related this with DNA methylation. The DNA composition is limited by only four nucleotides (DNA is made of just four nucleotides), while in RNA a great number of rare modified forms can be found. As is shown in works on the RNA biochemistry [17-19], nucleotide modification allows managing tautomerization in base pairs. In the context of these works, thymine can be considered as uracil with reduced capacity of tautomerization, which is justified by its function. N4-methoxycytosine is prone to form imino tautomer to such an extent that if forms a complementary pair with adenine [14]. Such a significant shift of the tautomer equilibrium would have become a source of mutations during replication and mistakes during transcription; therefor, the Nature chose methylation as a softer variant of the tautometry enhancement. 5-Methylcytosine has to possess an increased ability to form imino tautomer as compared with a common cytosine.

While in cytosine this ability of tautomerization was already high and was most expressed in the context of CpG, methylation of cytosine in this position may lead to an even greater increase of the proportion of tautomeres, non-canonic pairs, and, according to our supposition, to an increase in the frequency of the ultrasound-induced DNA cleavages.

A recent work of Al-Hashimi [20] sheds light on some questions concerning cytosine methylation. The contribution of cytosine methylation to the formation of the GC Hoogsteen Base Pair at pH from 5.5 to 7 was analyzed by means of the NMR methods. The pH range allows for estimation of the contribution of tautomere and protonated forms of cytosine. The main conclusion of the work is that there are no essential differences between C and MetC. Thus, Al-Hashimi in a series of works [8, 9, 11-12, 20] performed an in-depth analysis of the mechanisms and conditions of the emergence of the Transient Hoogsteen Base Pairs in the DNA structure and showed that methylation is notrelated with these pairs. As a result, Transient GC Wobble Base Pair remains the only possible molecular basis for the cytosine methylation effect upon ultrasound-induced DNA cleavage.

Let us specify the properties of the Transient GC Wobble Base Pair:

1. The propensity of deoxycytidine for the formation of a tautomer imino form favors the formation of this pair;

2. Being a structural analog of the GT Wobble Base Pair, a new pair creates a distortion in DNA, which can explain an increase in the frequency of the ultrasound-induces breakages of the sugar-phosphate backbone after deoxycytidine;

3. The AT pair is not able to form similar structures; this may explain an increase of the DNA cleavage frequency only after the deoxycytidine pairs;

4. GC Wobble Base Pair in the DNA) has to possess asymmetrical sugar-phosphate backbone; this explains an increase of the breakage frequency only after deoxycytidine;

5. Cytosine methylation is able to increase the stability of the GC Wobble Base Pair, that is, to increase the probability to meet such a Transient pair in the DNA composition. This explains an increase of the breakage frequency in dinucleotides CpG of the methylated DNA.

The assumption about the existence of the Transient GC Wobble Base Pair allows for explanation of experimental facts discovered in the course of studies of the ultrasound-induced DNA cleavage:

1. An increase of the frequency of the cleavage after deoxycytidine in all four dinucleotides due to the ability of C for tautometry and a possibility of the appearance of CG Wobble;

2. Asymmetry of the single-strand cleavages of DNA.

3. An increase of the frequency of the cleavage after cytosine methylation.

Methods developed by now make it possible to verify all our explanations. Among them are in the first place the NMR methods developed by Al-Hashimi for studies of Transient Base Pairs. GT Wobble Base Pair can become a model object in these investigations.

Methylation of deoxycytidine plays a very important role in epigenetics [21]. Transient GC Wobble Base Pair may also serve as an epigenetic marker. It is possible that methylation of deoxycytidine in DNA in the composition of CpG dinucleotides facilitates the appearance of the Transient GC Wobble Base Pair. Such pairs could be recognized by regulatory proteins. Obviously, the enzyme can find Transient GC Wobble Base Pair much easier than the GC pair with methylcytosine. Transient GC Wobble Base Pairs caused the distortion of the sugar-phosphate backbone and have two accessible amino groups capable of the formation of hydrogen bonds. Below we will indicate, where analogous structures may occur, and consider possibilities of further comparative analysis with an aim of elucidation of the mechanisms of the methylcytosine epigenetic effect. GU and GT Wobble Base Pair may appear in DNA as a result of the deamination mutation of cytosine and 5MetC, respectively. Reparation enzymes recognize these pairs, that is, they interact with the distortion in DNA and a free amino group. On the basis of the increased frequency of the ultrasound-induced cleavage of DNA we proposed a high frequency of the formation of Transient GC Wobble Base Pair in CpG dinucleotide. CpG is a target for three classes of proteins: cytosine(C5)-DNA-methyltransferase, DNA-demethylase, and regulatory proteins recognizing methylated cytosines in CpG in the DNA composition. Similarity of these three classes of proteins can be explained by structural similarity of the target substrates. We also predict similarity of these three classes of proteins with reparation enzymes of GT pairs on the basis of similarity of the GT Wobble Base Pair and Transient GC Wobble Base Pair.

ACKNOWLEDGMENTS

The authors thank Dr. Ilia V. Eltsov, who provided 1H15N NMR spectrum of deoxycytidine, Dr. Dmitry M. Graifer and Anastasia Anashkina, for useful discussion.

This work was supported program of the Presidium of the Russian Academy of Sciences for Molecular and Cellular Biology and the Program of Fundamental Research for State Academies for years 2013–2020, project no. 01201363818.